\documentclass[12pt]{iopart}

\usepackage{cite}
\usepackage{iopams}
\usepackage{subeqn}

\font\bigastfont=cmr10 scaled \magstep 1
\def\bdot{\hbox{\bigastfont .}}
\font\bigastfontt=cmr10 scaled \magstep 3
\def\bdott{\hbox{\bigastfontt .}}

\def\beq{\begin{equation}}
\def\eeq{\end{equation}}
\def\beqn{\begin{eqnarray}}
\def\eeqn{\end{eqnarray}}

\newcommand{\CD}{{\cal D}}

\newcommand{\CM}{{\cal M}}

\newcommand{\CQ}{{\cal Q}}
\newcommand{\CR}{{\cal R}}

\newcommand{\mnu}{{\mu \nu}}

\newcommand{\average}[1]{\left\langle #1 \right\rangle_\CD}
\newcommand{\averagew}[1]{\langle #1 \rangle_\CD}
\newcommand{\averageS}[1]{\left\langle #1 \right\rangle_{\Sigma}}
\newcommand{\averageSw}[1]{\langle #1 \rangle_{\Sigma}}

\newcommand{\initial}[1]{{#1_{\it i}}}

\newcommand{\wt}{\widetilde}

\newcommand{\leftexp}[2]{{\vphantom{#2}}^{#1}{#2}}

\begin{document}

\title[Relativistic cosmological perturbation scheme on a general background]
{Relativistic cosmological perturbation scheme on a general background: scalar perturbations for irrotational dust}

\author{Xavier Roy$^{1, 2}$$^{,\star}$ and Thomas Buchert$^{1}$}

\vspace{1mm}

\address{$^1$ Universit\'e de Lyon, Observatoire de Lyon, 
Centre de Recherche Astrophysique de Lyon, CNRS UMR 5574: Universit\'e Lyon~1 and \'Ecole Normale Sup\'erieure de Lyon, \\
9 avenue Charles Andr\'e, F--69230 Saint--Genis--Laval, France}

\address{$^2$ Astrophysics, Cosmology and Gravity Centre, Department of Mathematics and Applied Mathematics, University of Cape Town, 
Rondebosch 7701, South Africa}

\address{$^{\star}$ Claude Leon Postdoctoral Fellow

\medskip
Emails: xavier.roy@uct.ac.za, buchert@obs.univ-lyon1.fr}

\begin{abstract}
In standard perturbation approaches and $N$-body simulations, inhomogeneities are described to evolve on a predefined background 
cosmology, commonly taken as the homogeneous--isotropic solutions of Einstein's field equations (Friedmann--Lema\^itre--Robertson--Walker 
(FLRW) cosmologies). In order to make 
physical sense, this background cosmology must provide a reasonable description of the effective, i.e.\ spatially averaged, evolution of  
structure inhomogeneities also in the nonlinear regime. Guided by the insights that (i) the average over an inhomogeneous distribution 
of matter and geometry is in general not given by a homogeneous solution of general relativity, and that (ii) the class of FLRW cosmologies 
is not only locally but also globally gravitationally unstable in relevant cases, we here develop a perturbation approach that describes the 
evolution of inhomogeneities on a general background being defined by the spatially averaged evolution equations. This physical background 
interacts with the formation of structures. We derive and discuss the resulting perturbation scheme for the matter model `irrotational dust' in 
the Lagrangian picture, restricting our attention to scalar perturbations.
\end{abstract}

\pacs{98.80.-k, 95.36.+x, 98.80.Jk, 04.20.-q, 04.20.Cv, 04.25.Nx}




\section{A new approach to perturbation theory}

Perturbation theory is a key tool in cosmology to describe the formation of structures in the weakly nonlinear regime 
and to initialize the $N$-body simulations of cosmic structures. The standard motivation to describe perturbations on 
a homogeneous--isotropic Friedmann--Lema\^itre--Robertson--Walker (FLRW) background comes from the known 
gravitational instability of the latter. However, this background is also supposed, implicitly, to describe the average evolution 
of an inhomogeneous universe model in the nonlinear regime, and this is implemented as a construction principle in most known relativistic, 
quasi-Newtonian and Newtonian perturbation schemes and Newtonian $N$-body simulations \cite{buchert:jgrg}. 
We remark that such a construction is correct in Newtonian cosmology: if structures evolve on a Euclidean geometry, and if they are subjected to periodic 
boundary conditions on some large scale, then the average of inhomogeneities is given by the assumed background (see \cite{buchertehlers} for the proofs).

Obviously, a Newtonian cosmology has to be considered as highly restrictive when one moves to the framework of general 
relativity, and the previous construction cannot be expected to work there because of: (i) the relevance of the spatial intrinsic 
curvature (the second derivatives of the metric may be significant even if the metric perturbations are small \cite{buchert:estim,kolb:voids,kolb:backgrounds,kolb:focus,rasanen:perturbation,rasanenFOCUS}), 
together with (ii) the fact that inhomogeneities are coupled to the spatial curvature evolution \cite{buchert:fluid}, and finally 
(iii) the absence of a conservation law for the averaged intrinsic curvature \cite{buchertcarfora}. Also, it has been recently shown 
\cite{roy:ps}, in the special class of scaling laws for the spatially averaged inhomogeneities (the so-called backreaction terms), 
that FLRW backgrounds are not only locally but also 
globally unstable as a result of structure formation and accelerated expansion, when subjected to perturbations whose global contribution 
does not vanish. As soon as a homogeneous and isotropic self-gravitating system is perturbed, the related inhomogeneities 
invoke a departure of its average from a FLRW background. This property is of great significance for a theory of perturbations: 
in the standard approach the background evolution is known to impact on the evolution of perturbations, but the converse effect, 
namely that the structure inhomogeneities also affect the evolution of the background, has been neglected thus far by construction\footnote{Or it has been claimed to be a 
negligible effect. However, arguments for a small backreaction rely on a weakly perturbed FLRW 
cosmology, and only address the issue within a limited framework (see, e.g., the review \cite{clarkson:review} for a clear presentation of this issue, 
and references therein).}. Note that such a property is expected from first principles: it expresses the fact that, in Einstein's theory, the formation of structures 
and the evolution of geometry are mutually (and generically) coupled. An evolution of structures on a predefined background is, in 
light of these remarks, an {\em a priori} restricted approach.

Looking at perturbations in the universe, we can only apprehend their strength and their evolution correctly if we know with respect 
to which background they have to be considered. Previous work that has addressed the issue in Newtonian cosmology \cite{momoko1,momoko2} 
faces the drawback that, on some large scale, all averages are strictly free of backreaction from inhomogeneities due to the restriction to 
a non-dynamical geometry and the necessity of a torus architecture, as explained above. The same drawback remains in quasi-Newtonian relativistic 
perturbation schemes for the evolution of gauge-invariant variables on a predefined background, even if they take into account backreaction effects 
\cite{abramo1,morita,matsumoto,matarrese:pert,russ1,russ2,zhao}. 
A related discussion by R\"as\"anen can be found in \cite{rasanen:acceleration}. 
For example, the assumption of periodic boundary conditions on 
initially flat-space sections or the restriction of the scalar curvature to a constant curvature suppresses any interaction effect between structures and the 
background. Beyond the usual FLRW perturbation scheme, Clarkson {\it et~al}\ have furnished in \cite{clarkson:pert} a complete system of master 
equations that represent the general linear perturbations to Lema\^itre--Tolman--Bondi (LTB) cosmologies. In \cite{nakamura,giesel:pert}, a fully gauge-invariant 
relativistic perturbation theory has been given that holds for any background metric.

The framework presented in this paper provides the needed tools to implement a background as the average over fluctuating fields, which is 
not based on the introduction of a predefined background and deviations thereof; it furnishes the evolution of the scalar parts of the deviation fields off this 
general background in a non-perturbative way. We expect from this improvement that we shall be able to explicitly imprint structure inhomogeneities into 
the background that eventually describes the expansion history of the universe without the need to invoke, e.g., a dark energy fundamental component. 
A further aim is to understand initial conditions for relativistic numerical simulations of inhomogeneities that are not restricted to vanish on average 
on a FLRW background.

We proceed as follows. In section~\ref{sec:lac}, we briefly recall the local and averaged evolution and constraint equations for the description 
of inhomogeneous dust cosmologies, and we provide a system of equations for the deviation fields off a general background.
This is followed, in section~\ref{sec:discuss}, by a thorough discussion of the properties of this new deviation scheme. 
We conclude the paper in section~\ref{sec:conc} with some prospects.


\section{Perturbation scheme on a general background} \label{sec:lac}

\subsection{Local and averaged equations for inhomogeneous cosmologies}

Let us consider a globally hyperbolic four-dimensional manifold, endowed with some metric tensor $\leftexp{4}{\mathbf{g}}$. An irrotational fluid 
congruence, defined by a unit timelike vector field $\mathbf{u}$, will be used to foliate the spacetime into a family of flow-orthogonal space-like 
hypersurfaces. We shall restrict ourselves, in what follows, to the case of a pressureless irrotational  fluid (irrotational dust), 
$T_\mnu = \varrho u_\mu u_\nu$, with $\varrho$ its energy density, as described in the Lagrangian picture\footnote{Greek indices refer to 
spacetime components, they run in $\{0,1,2,3\}$, and Latin indices denote space components, running in $\{1,2,3\}$. 
For a presentation of the (3+1)-splitting of Einstein's equations, see e.g.\ \cite{alcub:fol,gourg:fol,smarr:fol}.}. 
In the canonical bases $(\partial_t, \partial_i)$ and $(dt, dX^i)$, the 4-velocity of the fluid assumes the form
\begin{subequations}

\beq
	u^\mu = (1, 0, 0, 0) \, , \qquad u_\mu = (-1,0,0,0) \, ,
\eeq
and the line element is written

\beq
	ds^2 = \leftexp{4}{g}_\mnu dX^\mu dX^\nu = - dt^2 + h_{ij} dX^i dX^j \, ,
\eeq
\end{subequations}
with $X^i$ being the Lagrangian spatial coordinates (coordinates comoving with the fluid), and $\mathbf{h}$ the inhomogeneous 3-metric 
of the $t$-constant hypersurfaces. The foliation of Einstein's equations with respect to $\mathbf{u}$ implies the well-known Raychaudhuri equation 
and Hamilton constraint:
\begin{subequations} \label{eq:rh}
\beqn
	&& \dot\Theta + \frac{1}{3}\Theta^2 = - 4 \pi G \varrho - 2 \sigma^2 + \Lambda \, , \label{eq:r} \\
	&& \frac{1}{3}\Theta^2 = 8 \pi G \varrho - \frac{\CR}{2} + \sigma^2 + \Lambda \, . \label{eq:h}
\eeqn
Throughout our study, the overdot will stand for the covariant derivative (here identical to the partial time derivative $\partial_t$). $\Theta := 
h^{ij} \Theta_{ij} := \frac{1}{2} h^{ij} \dot{h}_{ij}$ and $\sigma := (\frac{1}{2} \sigma^{ij} \sigma_{ij})^{1/2}$ are the local expansion and shear rates, respectively, 
while $\CR$ is the local three-Ricci scalar curvature of the hypersurfaces, and $\Lambda$ is the cosmological constant that we carry along for 
the sake of generality. Those relations may be supplemented by the fluid continuity equation and a balance relation between $\CR$ and $\sigma$:
\beqn
	&& \dot\varrho + \Theta \varrho = 0 \, , \label{eq:ce} \\
	&& \dot\CR + \frac{2}{3} \Theta \CR = 2 (\sigma^2)^{\bdot} + 4 \Theta \sigma^2 \, , \label{eq:ic}
\eeqn
\end{subequations}
where the latter is obtained upon requiring equation~(\ref{eq:h}) to be an integral of (\ref{eq:r}) (we shall simply call, hereafter, `integrability condition' the result of this procedure). The local system (\ref{eq:rh}) will be used, later on, to derive the evolution of the deviation fields inside any spatial domain. 
For notational ease, we have omitted the time and space dependences, but the reader should bear in mind that all variables at stake are inhomogeneous.

Note that for a locally isotropic cosmology, $\sigma = 0$, the system (\ref{eq:rh}) becomes homogeneous (Schur--Tr\"umper's theorem {\cite{schur,trump}}). 
We recover a FLRW cosmology. In particular, writing the expansion rate as $\Theta_H =: 3 \dot{a}_H/a_H$, with $a_H$ being the scale factor, the 
integrability condition (\ref{eq:ic}) implies that the spatial Ricci scalar curvature follows the evolution law: $\CR_H = 6k/a_H^2$, where $k$ is 
a constant of integration. 

Let us now consider a scalar field $\psi$. Its spatial average performed on some compact domain $\CD$, contained within the hypersurfaces and 
transported along the fluid flow lines (Lagrangian averaging), is defined as (see \cite{buchert:dust} for details)\footnote{Note that such a domain is `frozen' into the metric. 
It remains simply-connected for regular solutions, but it changes its morphology due to the time-dependence of the metric. 
Since the domain encloses during its evolution the same collection of fluid elements, there are no fluxes across its boundary, 
unlike in the corresponding Newtonian model where fluid elements move with respect to an external embedding space \cite{buchertehlers}.}:
\begin{subequations} 

\beq
	\average{\psi} := \frac{1}{V_\CD} \int_\CD \psi \sqrt{\det{h_{ij}}} \, d^3X \, ,
\eeq
with $V_\CD := \int_\CD \sqrt{\det{h_{ij}}} \, d^3X$ the volume of the domain under consideration satisfying $\dot{V}_\CD/V_\CD = \average\Theta$. 
We shall also make frequent use of the commutation rule between the spatial averaging and differentiation with respect to time:

\beq
	\average{\psi}^{\bdot} - \averagew{\dot\psi} = \average{\Theta \psi} - \average{\Theta} \average{\psi} \, ,
\eeq
\end{subequations}
where the right-hand side reduces to zero for a homogeneous domain. Equipped with these relations, we can provide the Lagrangian averaging 
on $\CD$ of Raychaudhuri's equation and Hamilton's constraint\footnote{For comprehensive reviews on averaged inhomogeneous cosmologies 
in general relativity, we recommend the reading of, e.g., \cite{rasanen:de,rasanen:acceleration,buchert:review,buchert:focus,clarkson:inho_conc,ellis:focus,wiltshire:focus,clarkson:review,buchertrasanen} 
(and references therein).}:
\begin{subequations} \label{eq:rh_av}
\beqn
	&&\average{\Theta}^{\bdot} + \frac{1}{3} \average\Theta^2 = - 4 \pi G \average\varrho + \CQ_\CD + \Lambda \, , \label{eq:r_av} \\
	&& \frac{1}{3} \average\Theta^2 = 8 \pi G \average\varrho - \frac{\average\CR}{2} - \frac{\CQ_\CD}{2} + \Lambda \, , \label{eq:h_av}
\eeqn
and we obtain the conservation law for the total rest mass within $\CD$ and the integrability condition for the averaged variables as
\beqn
	&& \average\varrho^{\bdot} + \average\Theta \average\varrho = 0 \, , \label{eq:ce_av} \\
	&& \average{\CR}^{\bdot} + \frac{2}{3} \average{\Theta} \average{\CR} + \dot{\CQ}_\CD + 2 \average{\Theta} \CQ_\CD = 0 \, , \label{eq:ic_av}
\eeqn
\end{subequations}
with $\CQ_\CD$ being the kinematical backreaction, 
\beq
	\CQ_\CD := \frac{2}{3} \average{(\Theta - \average\Theta)^2} - 2 \average{\sigma^2} \, . \label{eq:kb}
\eeq
Equations~(\ref{eq:rh_av}), the averaged counterpart of (\ref{eq:rh}), will also be used in the following to obtain the evolution of the deviation fields 
in the {\it interior} of $\CD$. $\CQ_\CD$ determines how the fluid inhomogeneities inside the domain globally contribute to the evolution of its background 
(equations (\ref{eq:r_av}) and (\ref{eq:h_av})), and this variable is dynamically coupled to the averaged scalar curvature (equation~(\ref{eq:ic_av})). 
Equation~(\ref{eq:ic_av}) also shows that the averaged curvature does not individually obey a conservation law like the fluid density; rather a combined 
expression of intrinsic and extrinsic curvature invariants is conserved\footnote{$\CQ_\CD$ can be written in terms of 
the extrinsic curvature of the hypersurfaces, $K_{ij} =: - \Theta_{ij}$, as 
$\CQ_\CD = \averagew{K^2 - K^{i j} K_{i j}} - \frac{2}{3} \average{K }^2 $.}. 
It is important to note that the 
background is scale-dependent: for another domain we have in general a different background, due to the unconstraint distribution of inhomogeneities. 
There so exists a deep correlation between the 
background of any domain and the inhomogeneities inside. This feature is habitually absent in the usual cosmological perturbation schemes on a global scale where perturbations are {\em assumed} to average out 
on a predefined background.

Note that if we continuously shrink the compact domain to a point (null-homotopy), 
$\average{(\Theta - \average\Theta)^2} \to 0$ and $\CQ_\CD \to - 2 \sigma^2$. 
The system~(\ref{eq:rh_av}) then reduces to~(\ref{eq:rh}). 

Finally, it is also convenient for later discussion to introduce two of the scalar invariants of the expansion tensor $\Theta^i_{\phantom{i} j} := 
\sigma^i_{\phantom{i} j} + \frac{1}{3} \Theta h^i_{\phantom{i} j}$, its trace and the dispersion of its non-diagonal components: 
\beq
	\mathrm{I} := \tr(\Theta^i_{\phantom{i} j}) = \Theta \, ,
	\qquad \mathrm{II} := \frac{1}{2} \left( \tr^2(\Theta^i_{\phantom{i} j}) - \tr(\Theta^i_{\phantom{i} j} \Theta^j_{\phantom{j} k}) \right) = \frac{1}{3} \Theta^2 - \sigma^2 \, . 
\eeq
The systems~(\ref{eq:rh}) and (\ref{eq:rh_av}) assume the same form with these variables \cite{buchert:dust}:
\begin{subequations}
\beqn
	&& \dot\mathrm{I} + \mathrm{I}^2 = 2 \mathrm{II} - 4 \pi G \varrho + \Lambda \, , \qquad \mathrm{II} = 8 \pi G \varrho - \frac{\CR}{2} + \Lambda \, , \label{eq:rh_si} \\
	&& \dot\CR + \frac{2}{3} \mathrm{I} \CR + \Big(2 \mathrm{II} - \frac{2}{3} \mathrm{I}^2 \Big)^{\bdot} + 2 \mathrm{I} \Big(2 \mathrm{II} - \frac{2}{3} \mathrm{I}^2 \Big) = 0 \, , \label{eq:icl}
\eeqn
for the local one, and 
\beqn
	&& \hspace{-1.5cm} \average{\mathrm{I}}^{\bdot} + \average{\mathrm{I}}^2 = 2 \average{\mathrm{II}} - 4 \pi G \average\varrho + \Lambda \, , \qquad \average{\mathrm{II}} = 8 \pi G \average\varrho - \frac{\average\CR}{2} + \Lambda \, , \label{eq:rh_si_av} \\
	&& \hspace{-1.5cm} \average\CR^{\bdot} + \frac{2}{3} \average{\mathrm{I}} \average\CR + \Big(2 \average{\mathrm{II}} - \frac{2}{3} \average{\mathrm{I}}^2 \Big)^{\bdot} + 2 \average{\mathrm{I}} \Big(2 \average{\mathrm{II}} - \frac{2}{3} \average{\mathrm{I}}^2 \Big) = 0 \, , \label{eq:ic2}
\eeqn
\end{subequations}
for the averaged one. We have reformulated the kinematical backreaction as $\CQ_\CD = 2 \average{\mathrm{II}} - \frac{2}{3} \average{\mathrm{I}}^2$ to obtain 
the second set of expressions.
This statement obviously holds true for the local and averaged continuity equations.

\subsection{Equations for the deviation fields}

We now provide the whole set of equations for the deviation fields off the background of a comoving dust domain. As is customary, we shall designate the deviation 
(or {\it peculiar}) field of any scalar field $\psi$ from its background value by $\delta\psi := \psi - \average\psi$. 
One would prefer to write $\delta_\CD\psi$ in order to make explicit the scale-dependence of the deviations; however, 
we drop this index for notational ease. Note finally that, $\average\psi$ being a scalar (refer to \cite{gasp} for a proof), the deviation $\delta\psi$ is also a scalar. 

In this paragraph, we only add a few remarks about each proposition. A thorough discussion follows in section~\ref{sec:discuss}. 


\subsubsection{Deviations in density.}

Using the local and averaged conservation laws~(\ref{eq:ce}) and (\ref{eq:ce_av}), we find the following continuity equation for the fluid density 
deviations, which we formulate in the form of a first proposition.

\bigskip
\noindent
{\bf Proposition 1a.}
{\it The evolution equation for the density deviations on a compact domain $\CD$ is given by}
\begin{subequations} \label{eq:pert_sys_1}
\beq
	(\delta\varrho)^{\bdot} + \average\Theta \delta\varrho = - \delta\Theta \, \left( \average\varrho + \delta\varrho \right) \, , \label{eq:ce_pert}
\eeq
{\it or, equivalently, in terms of scalar invariants by}
\beq
	(\delta\varrho)^{\bdot} + \average{\mathrm{I}} \delta\varrho = - \delta\mathrm{I} \, \left( \average\varrho + \delta\varrho \right) \, .
\eeq

\bigskip
\noindent
{\bf Remark.}
The density deviation field does not obey a conservation law like the local and averaged densities. We are faced with a source term involving 
the deviation of the expansion rate from its background value. By making use of the commutation rule, the average on $\CD$ of these equations 
results in identities.

\bigskip
\noindent
For later discussion we shall prefer to use an alternative form of this proposition. Consider to this end the scale-dependent contrast density 
$\Delta_\CD := \delta\varrho/\varrho$ ($-\infty < \Delta_\CD < 1$), which is more adapted to the Lagrangian picture and the nonlinear situation 
\cite{buchert89, buchert92,buchert:nonperturbative}, than the conventional definition used in Eulerian perturbation theory, 
$\delta_\CD := \delta\varrho/\average\varrho$ (restricted to the same domain). 
By means of the local and averaged Raychaudhuri equations~(\ref{eq:r}) and (\ref{eq:r_av}), we end up with the following evolution equations for 
$\Delta_\CD$.

\bigskip
\noindent
{\bf Proposition 1b.} 
{\it The evolution equations for the contrast density on a compact domain $\CD$ are written as}
\beqn
	&& \dot{\Delta}_\CD = \delta\Theta \, (\Delta_\CD - 1) \, , \label{eq:fod} \\
	&& \ddot{\Delta}_\CD + \frac{2}{3} \average\Theta \dot{\Delta}_\CD - 4 \pi G \average\varrho \Delta_\CD = \delta\CQ \, (\Delta_\CD - 1) \, , \label{eq:sod}
\eeqn
{\it or, equivalently, in terms of scalar invariants as}
\beqn
	&& \dot{\Delta}_\CD = \delta\mathrm{I} \, (\Delta_\CD - 1) \, , \\
	&& \ddot{\Delta}_\CD + \frac{2}{3} \average{\mathrm{I}} \dot{\Delta}_\CD - 4 \pi G \average\varrho \Delta_\CD = 2 \left( \delta\mathrm{II} - \frac{2}{3} \average{\mathrm{I}} \delta\mathrm{I} \right) (\Delta_\CD - 1) \, .
\eeqn
\end{subequations}

\bigskip
\noindent
{\bf Remark.}
We have introduced here the local contribution of fluid inhomogeneities within the domain,  
$\CQ := \frac{2}{3} (\Theta - \average\Theta)^2 - 2 \sigma^2 = 2 \mathrm{II} - \frac{2}{3} \average{\mathrm{I}} ( 2 \mathrm{I} - \average{\mathrm{I}})$. 
By construction we have $\average\CQ = \CQ_\CD$, and $\delta\CQ$ stands for the deviation of the kinematical backreaction. Taking the averages of these 
relations and using the definition of $\Delta_\CD$  and the commutation rule, we obtain identities, as it should be for a proper definition of deviation fields.


\subsubsection{Deviations in kinematical variables.}

Using the local and averaged systems, ((\ref{eq:r}) and (\ref{eq:h})) and ((\ref{eq:r_av}) and (\ref{eq:h_av})) respectively, 
we find the following equations for the kinematical deviations. 

\bigskip
\noindent
{\bf Proposition 2.}
{\it The evolution and constraint equations for the kinematical deviations on a compact domain $\CD$ read}
\begin{subequations} \label{eq:pert_sys_2}
\beqn
	&& (\delta\Theta)^{\bdot} + (\delta\Theta)^2 + \frac{2}{3} \average\Theta \delta\Theta 
	= - 4 \pi G \delta\varrho + \delta\CQ \, , \label{eq:r_pert} \\
	&& \frac{2}{3} \average\Theta \delta\Theta 
	= 8 \pi G \delta\varrho  - \frac{1}{2} \delta\CR - \frac{1}{2} \delta\CQ \, , \label{eq:h_pert}
\eeqn
{\it or, equivalently, in terms of scalar invariants} 
\beqn
	&& (\delta\mathrm{I})^{\bdot} + (\delta\mathrm{I})^2 + 2 \average{\mathrm{I}} \delta\mathrm{I} 
	= 2 \, \delta\mathrm{II} - 4 \pi G \delta\varrho \, , \label{eq:r_pert2} \\
	&& \delta\mathrm{II} = 8 \pi G \delta\varrho - \frac{1}{2} \delta\CR \, . \label{eq:h_pert2}
\eeqn
\end{subequations}

\bigskip
\noindent
{\bf Remark.}
Contrary to the relation~(\ref{eq:h_pert2}), the shape of (\ref{eq:r_pert2}) is not identical to that of its local and averaged counterparts, 
(\ref{eq:rh_si}) and (\ref{eq:rh_si_av}). The nonlinear character of the latter makes the extra term $2 \average{\mathrm{I}} \delta\mathrm{I}$ appear. 
Taking the averages of (\ref{eq:r_pert2}) and (\ref{eq:h_pert2}), and using the commutation rule for the first one, we end up with identities. 


\subsubsection{Integrability condition.}

Finally, demanding equation~(\ref{eq:h_pert}) to be an integral of (\ref{eq:r_pert}), we obtain the integrability condition between the kinematical 
deviations and the intrinsic curvature, which we formulate in the form of a last proposition.

\bigskip
\noindent
{\bf Proposition 3.}
{\it The integrability condition on a compact domain $\CD$ reads}
\begin{subequations} \label{eq:pert_sys_3}
\beq
	\hspace{-2cm} (\delta\CR)^{\bdot} + \frac{2}{3} \average\Theta \delta\CR + (\delta\CQ)^{\bdot} + 2 \average\Theta \delta\CQ 
	= - \delta\Theta \left( \frac{2}{3} \average\CR + 2 \CQ_\CD + \delta\CR + \delta\CQ \right) \, , \label{eq:ic_pert}
\eeq
{\it or, equivalently, in terms of scalar invariants}
\beq
	(\delta\CR)^{\bdot} + \average{\mathrm{I}} \delta\CR + 2 (\delta\mathrm{II})^{\bdot} + 2 \average{\mathrm{I}} \delta\mathrm{II}
	= - 16 \pi G \, \delta\mathrm{I} \, \left( \average\varrho + \delta\varrho \right) \, .
\eeq
\end{subequations}

\bigskip
\noindent
{\bf Remark.}
The curvature and backreaction deviations are coupled through this integrability condition. 
Here again a source term involving the deviation of the expansion rate is present; in terms of scalar invariants, this term is the same as that of 
proposition 1a. The average of these relations also results in identities. 


\section{Properties of the deviation scheme} \label{sec:discuss}

\subsection{Discussion}

We have generalized the usual dust perturbation scheme off a predefined background to deviations off a general background that was obtained 
through the spatial average of inhomogeneous fields on a generic domain. The background of a pressureless self-gravitating system is then not 
restricted to follow a predefined evolution, but rather an evolution depending on the inhomogeneous distribution of matter and geometry. 
As we shall see further below, deviations off FLRW backgrounds are recovered for globally isotropic domains having a vanishing kinematical backreaction; 
they constitute a subclass of solutions within the present framework. Here, we prefer to speak in terms of deviations rather than perturbations since, by 
construction, this scheme is non-perturbative (no approximation or linearization has been performed). A consequence of this property, and accordingly 
a second interesting feature of this approach, is that we do not get any constraint on the strength of deviations, apart from the requirement of regularity 
of the solutions\footnote{Regularity is violated in the presence of shell-crossing singularities. Shell-crossing happens
generically due to the chosen matter model `irrotational dust', since dispersion and vorticity are not taken into account
to regularize the solutions (for generalizations and discussions in the Newtonian theory, see \cite{buchertdominguez:adhesive}).
The inclusion of backreaction effects improves on this situation in so far as large perturbations on a FLRW background would be
eventually mirrored by small perturbations on the general background, so that the scale of the perturbations undergoing shell-crossing could be smaller. 
Note also that for volume averages on sufficiently large scales the inclusion of dispersion and vorticity is not expected to be quantitatively
relevant, since those effects are sizable within small fractions of the volume only.}. 

It is also worth noting that this scheme only functionally depends on a metric (via the domain of averaging); 
all the equations outlined above keep the same form for any spatial metric. The 
role of the 3-metric was indeed entirely implicit for the derivation of our scheme: we did not need to compute an averaged metric from the local one in 
order to eventually obtain the dynamics of the deviation fields off a general background. This is a nice feature, and one can choose any spatial metric and 
end up with a solution. We recognize that the disadvantage is to be able to deal only with the scalar modes of the deviations, and not the vectorial and 
tensorial ones. However, we think that the results expounded here clearly constitute a first useful step for an understanding of cosmological deviations off a general 
background.

The generalization we have proposed shows that the kinematical backreaction, which encodes the global contribution of fluid inhomogeneities within a 
spatial domain, impacts in several ways on the dynamics of the deviations: not only $\CQ_\CD$ affects the evolution of deviations through that of the 
background, as it is implied by the coefficients like $\average{\Theta}$ on the left-hand side of equations~(\ref{eq:sod}), (\ref{eq:r_pert}) and (\ref{eq:h_pert}), 
but it also acts as a source through the peculiar-backreaction $\delta\CQ$ (right-hand side of~(\ref{eq:sod}), (\ref{eq:r_pert}) and (\ref{eq:h_pert})). 
Comparing the evolution of averaged inhomogeneous cosmologies to that of (scale-dependent) FLRW models, it appears natural that the extra term involved 
in the evolution of density deviations off a general background is precisely the kinematical backreaction (see subsection~\ref{subsec:limit}). We also stress
the difference between a general domain and a FLRW domain: the spatial curvature of the former is in general not given by a 
constant-curvature model. The different evolution histories of the background curvature influence the dynamics of the background, which modifies in 
turn the evolution of the deviation fields---again in comparison to that of a FLRW background. In addition, inhomogeneities in geometry also act as a source 
in the evolution equations (the term $\delta\CR$ in equation~(\ref{eq:h_pert})).

\subsection{Steps toward an exact resolution} \label{subsec:exact_resol}
 
Let us first consider the systems for the local variables, the averaged variables and the deviation variables individually.
\begin{itemize}
	\item The local system~(\ref{eq:rh}) needs one additional relation to be solved, since we deal with three independent expressions for four variables 
	($\varrho$, $\Theta$, $\sigma$ and $\CR$). The hierarchy would continue with the evolution equation for the shear, but it will never be closed on the 
	level of ordinary differential equations (see \cite{buchert:dust} and also \cite{kofman}).
	\item The averaged system~(\ref{eq:rh_av}) also requires a last relation (one on the averaged variables) since in this situation we also have three 
	independent expressions for four variables ($\average\varrho$, $\average\Theta$, $\CQ_\CD$ and $\average\CR$) (see ibid.).
	\item Finally, the deviation field system~((\ref{eq:pert_sys_1}),(\ref{eq:pert_sys_2}) and (\ref{eq:pert_sys_3})) calls for three additional relations, 
	since we have three independent 
	equations for now six variables ($\delta\varrho$, $\average\varrho$, $\delta\Theta$, $\average\Theta$, $\delta\CQ$ and $\delta\CR$).
\end{itemize}
However, the two last systems are obviously related to the first one. Given any closure relation for the local system, we can solve all sets of equations 
and eventually obtain an exact solution for the deviation scheme. One can think of choosing a specific form for the spatial metric, 
or of giving a constraint on kinematical or geometrical variables (e.g.\ considering a `silent universe model' \cite{bruni}). 
For instance, considering a domain endowed with a spatial LTB metric, we are able to construct its background and compute its averaged evolution. 
According to our scheme, deviations from this background have to be understood as local perturbations giving back the local LTB metric. This can be applied 
to any synchronous metric. Although maybe not suited to furnish relevant physical models, this first procedure may provide interesting toy models. 

If we instead specify a constraint on the averaged variables on 
a chosen scale, we cannot end up with a solution for the deviation scheme. Nevertheless, by doing so we reduce the needed number of extra relations 
to one in order to exactly resolve it, since there will remain four variables ($\delta\varrho$, $\delta\Theta$, $\delta\CQ$ and $\delta\CR$) for the three-equation 
system~((\ref{eq:pert_sys_1}),(\ref{eq:pert_sys_2}) and (\ref{eq:pert_sys_3})). 
The second approach to obtain a solution may then be realized by reducing the space of possible backgrounds, e.g.\ with 
scaling laws \cite{roy:ps,buchert:morphon}, or particular effective state equations for the backreaction terms \cite{roy:chap,buchert:inf}, or multi-scale 
partitionings combined with closure assumptions, e.g. \cite{wiltshire:cosmic,wiltshire:average,multiscale}, and then by considering the resulting equations 
for the deviation fields. 
Although being straightforward to get working models of structure formation, this latter approach shall always call for physical verification of the closure relations used.

\subsection{Definition of a global physical background and deviations thereof} \label{subsec:pb}

We dedicate this subsection to the reformulation of the deviation field system~((\ref{eq:pert_sys_1}),(\ref{eq:pert_sys_2}) and (\ref{eq:pert_sys_3})) 
with the help of a spatial metric comoving with 
the global physical background, that we shall define in the ensuing paragraph\footnote{The formalism propounded hereafter may as well be viewed 
as a reformulation of our scheme for spatial coordinates rescaled by the scale factor of the global background.
Although reminiscent of the standard procedure of introducing `comoving coordinates', we shall not pursue this possibility since our 
approach is coordinate-invariant in the hypersurfaces, and such coordinate changes would add no physical insight here. Comoving coordinates 
make sense if a global coordinate system, e.g., on a constant-curvature domain, can be introduced. In general this is not possible, and a conformal 
transformation of local coordinates seems unnecessary (it may be of technical help in calculations).}.


\subsubsection{The global physical background in cosmology.}

We have discussed thus far the evolution and constraint equations for the deviation fields off a general physical background, 
obtained from the spatial averaging procedure. 
Let us now consider a compact spatial domain of the universe---we shall call it $\Sigma$---that we may assume {\it to cover the homogeneity 
scale}, namely that spatial scale beyond which all averages do no longer depend on scale. Such an assumption, common 
to cosmology, is not necessary, but it enables us to define a scale over which we think the universe is representative (any larger scale would not 
provide new insights) and to have a more transparent frame of comparison with the standard cosmological model and its usual perturbation schemes. 
We emphasize, again, that the following reformulation does not rely on the existence of a homogeneity scale, and it may equivalently be employed 
to describe globally inhomogeneous models, in which case $\Sigma$ would cover the whole spatial manifold, e.g.\ a spherical space without boundary (see 
\cite{buchert:static} for the average properties of such universe models).

We shall identify the scale of homogeneity as the one where the global physical background is defined (in practice such a scale would correspond 
to statistical homogeneity---see \cite{labini:focus} for an examination of the subject). The idea realized here corresponds to the `average background solution' 
discussed by Kolb and collaborators \cite{kolb:backgrounds,kolb:focus}. All averages indexed by $\Sigma$ shall then refer to `global' 
averages that define our background.


\subsubsection{A globally volume-preserving metric.}

Let us introduce, on $\Sigma$, the conformally rescaled Riemannian 3-metric $\mathbf{\wt{h}}$ as
\begin{subequations}
\beq
	h_{ij} :=  a_{\Sigma}^2 \, \widetilde{h}_{ij} \, ,
	\qquad h^{ij} := a_{\Sigma}^{-2} \, \widetilde{h}^{ij} \, ,
	\qquad h^i_{\phantom{i} j} = \widetilde{h}^i_{\phantom{i} j} = \delta^i_{\phantom{i} j} \, , \label{eq:cm}
\eeq
where $a_\Sigma$ is the dimensionless effective scale factor of the global domain $\Sigma$, defined as, and satisfying:
\beq
	a_\Sigma := \left( \frac{V_\Sigma}{V_{\Sigma_i}} \right)^{1/3} \, ,
	\qquad 3 \frac{\dot{a}_\Sigma}{a_\Sigma} = \frac{\dot{V}_\Sigma}{V_\Sigma} = \averageS{\Theta} \, , \label{eq:sf_s}
\eeq
\end{subequations}
with $\Sigma_i$ being the domain at initial time. The definition~(\ref{eq:cm}) guarantees that the metric 
$\mathbf{\wt{h}}$ conserves the volume of the domain $\Sigma$: $\wt{V}_\Sigma := \int_\Sigma (\det{\widetilde{h}_{ij}})^{1/2} \, d^3X = a_\Sigma^{-3} V_\Sigma$. 
We may say that $\mathbf{\wt{h}}$ stands for the spatial metric `comoving' with the global background. Remark that, the conformal 3-metric being 
inhomogeneous, we still have local variations of volume elements in $\mathbf{\wt{h}}$.

We now propose to construct, in this metric, all the scalar fields we shall need for the reformulation of our scheme. Upon defining the corresponding 
expansion tensor as
\begin{subequations}

\beq
	\wt{\Theta}^i_{\phantom{i} j} := \frac{1}{2} \widetilde{h}^{ik} \dot{\widetilde{h}}_{kj} \, ,
\eeq
we are able to write, with the help of expressions~(\ref{eq:cm}) and (\ref{eq:sf_s}):

\beq
	\wt\Theta^i_{\phantom{i} j} = \Theta^i_{\phantom{i} j} - \frac{1}{3} \averageS{\Theta} h^i_{\phantom{i} j} 
	= \sigma^i_{\phantom{i} j} + \frac{1}{3} (\Theta - \averageS{\Theta}) \, h^i_{\phantom{i} j} \, . \label{eq:tet}
\eeq
\noindent
The trace of this equality and the average of the resulting expression, respectively, yield

\beq
	\wt\Theta = \Theta - \averageS\Theta = \delta\Theta \, , \qquad \averageSw{\wt{\Theta}} = 0 \, . \label{eq:perS}
\eeq
\end{subequations}

\noindent
The first relation reveals that $\wt\Theta$ pinpoints the deviation of the local expansion rate from the $\mathbf{h}$-background expansion rate 
of the global domain, and it thus defines the peculiar-expansion rate of the latter. These expressions are consistent with the stationarity of the 
$\mathbf{\wt{h}}$-background of $\Sigma$, $\averageSw{\wt{\Theta}} = 0$, and with the existence, in general, of local variations of volume 
elements evaluated with $\mathbf{\wt{h}}$, $\wt\Theta \neq 0$ ($\wt\Theta$ vanishes only if the global domain is homogeneous).

By means of equation~(\ref{eq:tet}), the traceless part of the `tilde' expansion tensor is the `tilde' shear tensor of the fluid defined as 
$\wt\sigma^i_{\phantom{i} j} := \sigma^i_{\phantom{i} j}$, which implies $\widetilde{\sigma}^2 = \sigma^2$. 
Concerning the spatial curvature of the global domain, a straightforward calculation, calling for the use of 
equation~(\ref{eq:cm}), results in $\wt\CR = a^2_\Sigma \CR$. 

Finally, the tilde energy density $\wt\varrho$ is obtained by considering the fluid conservation law
\begin{subequations}

\beq
	\dot{\widetilde\varrho} + \widetilde{\Theta} \, \widetilde\varrho = 0 \, ,
\eeq
which is simply the counterpart of (\ref{eq:ce}) in the spatial geometry generated from the conformally rescaled metric. Writing

\beq
	\widetilde{\Theta} = \Theta - \averageS\Theta = - \frac{\dot\varrho}{\varrho} + \frac{\averageS\varrho^{\bdot}}{\averageS\varrho} 
	= - \left( \frac{\varrho}{\averageS\varrho} \right)^{\bdot} \, \frac{\averageS\varrho}{\varrho} \, ,
\eeq
where we have used the local and averaged conservation laws for the second equality, we obtain

\beq
	\left( \frac{\varrho}{\averageS\varrho} \right)^{\bdot} + \widetilde{\Theta} \, \frac{\varrho}{\averageS\varrho} = 0 \, .
\eeq

\noindent
Multiplying both sides by the initial averaged density, we conclude

\beq
	\wt\varrho := \frac{\varrho}{\averageS\varrho} \left\langle \varrho \right\rangle_{\Sigma_{\it i}} \, ,
	\qquad \averageSw{\wt\varrho} = \left\langle \varrho \right\rangle_{\Sigma_{\it i}} = const. 
\eeq
\end{subequations}

\noindent
The last relation is naturally expected: the total rest mass within the global domain being conserved, the matter tilde density has to remain 
{\it globally} constant in the frozen volume $\wt{V}_\Sigma$. (Consider the invariant total rest mass 
$\CM = \wt\CM = \wt{V}_\Sigma \averageS{\wt\varrho}$; its conservation indeed implies $\averageS{\wt\varrho}^{\bdot} = 0$.) We also remark that 
$\sigma$ is the only scalar field we shall use that is not affected by the conformal rescaling. These characteristics are due to the fact 
that $\mathbf{\wt{h}}$ is built such to provide a volume-preserving metric and thus only transforms quantities related to the trace part of tensors.

In terms of scalar invariants, we are able to write the following equalities:
\begin{subequations}

\beq
	\mathrm{I} = \averageS{\mathrm{I}} + \mathrm{\wt{I}} \, ,
	\qquad \mathrm{II} = \mathrm{\wt{II}} + \frac{1}{3} \averageS{\mathrm{I}}^2 + \frac{2}{3} \averageS{\mathrm{I}} \mathrm{\wt{I}} \, ,
	\label{eq:si_tilde}
\eeq
where we have introduced the scalar invariants of $\wt\Theta^i_{\phantom{i} j}$ as 
$\mathrm{\wt{I}} := \tr(\wt\Theta^i_{\phantom{i} j}) = \wt\Theta$ and 
$\mathrm{\wt{II}} := \frac{1}{2} ( \tr^2(\wt\Theta^i_{\phantom{i} j}) - \tr(\wt\Theta^i_{\phantom{i} j} \wt\Theta^j_{\phantom{j} k}) ) 
= \frac{1}{3} \wt\Theta^2 - \wt\sigma^2$. Averaging over $\Sigma$ the last relation, and inserting the result back into it, gives\footnote{To be 
rigorous we should write $\wr\, \wt\psi \,\wr_\Sigma$ the average over $\Sigma$ of any tilde scalar field, with 
$\wr \bdott \wr_\Sigma := (1/ \wt{V}_\Sigma) \int_\Sigma \bdott \sqrt{\det{\wt{h}_{ij}}} \, d^3X$. However, one can easily check that, 
for any spacetime scalar field, $\wr \bdott \wr_\Sigma = \averageS{\bdott}$.}

\beq
	\mathrm{II} = \averageS{\mathrm{II}} +  \mathrm{\wt{II}} - \averageSw{\mathrm{\wt{II}}} + \frac{2}{3} \averageS{\mathrm{I}} \mathrm{\wt{I}} \, .
	\label{eq:si_tilde_2}
\eeq
\end{subequations}
$\mathrm{\wt{I}} = \delta\mathrm{I}$ represents the deviation field of the first local scalar invariant from its average on the global domain, 
and $\mathrm{\wt{II}} - \averageSw{\mathrm{\wt{II}}} + \frac{2}{3} \averageS{\mathrm{I}} \mathrm{\wt{I}} = \delta\mathrm{II}$ 
is that of the second local scalar invariant. Using these relations, we also recast the local contribution of fluid inhomogeneities into 
$\wt\CQ = \frac{2}{3} \wt\Theta^2 - 2 \wt\sigma^2 = 2 \mathrm{\wt{II}} = \CQ$, and the global one into 
$\wt\CQ_\Sigma = \frac{2}{3} \averageSw{\wt\Theta^2} - 2 \averageSw{\wt\sigma^2} = 2 \averageSw{\mathrm{\wt{II}}} = \CQ_\Sigma$ 
(therefore $\delta\CQ = \delta\wt\CQ = \delta\mathrm{\wt{II}}$).


\subsubsection{Deviations off a global physical background.} \label{subsec:deviationsSigma}

From the propositions given in section~\ref{sec:lac} a set of corollaries follows. It determines the evolution and constraint equations for the 
deviation fields off the global background, as expressed in a globally volume-preserving metric. From this point of view, deviations do not `see' 
and are not affected by the volume deformation of the background. We write the system in terms of the tilde scalar invariants, which can be straightforwardly found by using the above definitions. 

\bigskip
\noindent
{\bf Corollary 1.}
{\it The evolution equations for the contrast density on a global background $\Sigma$ read}
\begin{subequations}
\beqn
	&& \dot{\wt\Delta}_\Sigma = \mathrm{\wt{I}} \, (\wt\Delta_\Sigma - 1) \, , \\
	&& \ddot{\wt\Delta}_\Sigma + 2 \, \frac{\dot{a}_\Sigma}{a_\Sigma} \, \dot{\wt\Delta}_\Sigma - 4 \pi G \frac{\averageSw{\wt\varrho}}{a^3_\Sigma} \wt\Delta_\Sigma 
	= 2 \, \delta\mathrm{\wt{II}} \, (\wt\Delta_\Sigma - 1) \, . \label{eq:sod_tilde}
\eeqn
\end{subequations}

\bigskip
\noindent
{\bf Remark.}
We have used the equality $\Delta_\Sigma = \delta\varrho/\varrho = \delta\wt\varrho/\wt\varrho = \wt\Delta_\Sigma$ 
and definition (\ref{eq:sf_s}).

\bigskip
\noindent
{\bf Corollary 2.}
{\it The evolution and constraint equations for the tilde scalar invariants on a global background $\Sigma$ are written as}
\begin{subequations}
\beqn
	&& \dot{\mathrm{\wt{I}}} + \mathrm{\wt{I}}^{\,2} + 2 \, \frac{\dot{a}_\Sigma}{a_\Sigma} \, \mathrm{\wt{I}} 
	= 2 \, \delta\mathrm{\wt{II}} - 4 \pi G \frac{\delta\wt\varrho}{a^3_\Sigma} \, , \\
	&& \delta\mathrm{\wt{II}} + 2 \, \frac{\dot{a}_\Sigma}{a_\Sigma} \, \mathrm{\wt{I}} 
	= 8 \pi G \frac{\delta\wt\varrho}{a^3_\Sigma} - \frac{1}{2} \frac{\delta\wt\CR}{a^{2}_\Sigma} \, .
\eeqn
\end{subequations}

\bigskip
\noindent
{\bf Corollary 3.}
{\it The integrability condition on a global background $\Sigma$ is given by}

\beq
	\frac{(\delta\wt\CR)^{\bdot}}{a^{2}_\Sigma} + 2 (\delta\mathrm{\wt{II}})^{\bdot} + 12 \, \frac{\dot{a}_\Sigma}{a_\Sigma} \, \delta\mathrm{\wt{II}} 
	= - \mathrm{\wt{I}} \left( \frac{2}{3} \frac{\averageSw{\wt\CR}}{a^{2}_\Sigma} + 4 \averageSw{\mathrm{\wt{II}}} 
	+ \frac{\delta\wt\CR}{a^{2}_\Sigma} + 2 \, \delta\mathrm{\wt{II}} \right) \, .
\eeq

\subsection{The limit of a FLRW background} \label{subsec:limit}

FLRW cosmologies are recovered only if, for any compact region $\CD$ lying in the interior of $\Sigma$, we ask for the vanishing of $\CQ_\CD$ 
(the integrability condition~(\ref{eq:ic_av}) then imposes $\CR$ to be the Friedmannian curvature). In this situation, any scalar field equals 
its background value; the local and averaged systems~(\ref{eq:rh}) and (\ref{eq:rh_av}) are identical, and there do not exist deviations over $\Sigma$. 
The existence of deviations, and hence the possibility to form structures, then demands the global domain not to remain locally isotropic. In other words, 
we need to abandon the strong cosmological principle in favor of, for instance, a weaker version that defines a scale of homogeneity 
(see subsection~\ref{subsec:pb}). We may for example consider $\Sigma$ to follow globally, and not locally, a 
FLRW evolution, requiring the kinematical backreaction to vanish on the homogeneity scale $\Sigma$, and therefore on any larger scale. Such an assumption 
ensures that the background is globally Friedmannian, but at the same time does not prevent local inhomogeneities to live inside the global domain. 
In this picture, the cancellation of $\CQ_\Sigma$ would be the result of an exact compensation between the expansion variance and the average 
of the shear squared. (This is what happens, for instance, for spatially averaged zero-curvature LTB models, see \cite{buchert:focus} for the proofs and e.g.\ 
\cite{LTB:review,sussman:ltb} for further details.)

Facing such a global domain constrains the averaged system with the condition $\CQ_\Sigma = 2\averageSw{\mathrm{\wt{II}}} = 0$, and 
allows us to write the second-order differential equation of the contrast density (\ref{eq:sod_tilde}) as follows:
\begin{subequations} \label{eq:dev_flrw}

\beq
	\ddot{\wt\Delta}_\Sigma + 2 \frac{\dot{a}_\Sigma}{a_\Sigma} \dot{\wt\Delta}_\Sigma - 4 \pi G \frac{\averageSw{\wt\varrho}}{a^3_\Sigma} \wt\Delta_\Sigma 
	= 2 \mathrm{\wt{II}} \, (\wt\Delta_\Sigma - 1) \, , \label{eq:delta}
\eeq
where $a_\Sigma$ is the scale factor of the background satisfying

\beq
	3 \frac{\ddot{a}_\Sigma}{a_\Sigma} = - 4 \pi G \frac{\langle \varrho \rangle_\initial\Sigma}{a_\Sigma^{3}} + \Lambda \, , 
	\qquad 3 \left( \frac{\dot{a}_\Sigma}{a_\Sigma} \right)^2 = 8 \pi G \frac{\langle \varrho \rangle_\initial\Sigma}{a_\Sigma^{3}} 
	- \frac{1}{2} \frac{\langle \CR \rangle_\initial\Sigma}{a_\Sigma^{2}} + \Lambda \, .
\eeq
\end{subequations}
We recover with~(\ref{eq:dev_flrw})  the standard framework of density deviations off a FLRW background cosmology \cite{buchert:perturb}, 
which therefore constitutes a subclass of solutions of our deviation scheme. We leave it to the reader to simplify the other deviation field equations 
(corollaries 2 and 3) using the above condition ($\delta\mathrm{\wt{II}} = \mathrm{\wt{II}}$). 
Let us further suppose that the fluid inhomogeneities weakly contribute to the local kinematics of the deviation fields: $\CQ = \wt\CQ \approx 0$. 
We can then neglect the quadratic invariant $\mathrm{\wt{II}}$ and end up with
\begin{subequations}
\beq
	\ddot{\wt\Delta}_\Sigma + 2 \frac{\dot{a}_\Sigma}{a_\Sigma} \dot{\wt\Delta}_\Sigma - 4 \pi G \frac{\averageSw{\wt\varrho}}{a^3_\Sigma} = 0 \, , \label{eq:zeld} 
\eeq
which gives the evolution of the first-order Lagrangian (relativistic) density perturbations off a FLRW background. The same expression is obtained for 
the evolution of the linear density perturbations in standard perturbation theory. This equation corresponds to the linearization in $\wt\Delta_\Sigma$ of 
(\ref{eq:sod_tilde}), and it is solved by the exact relativistic form of Zel'dovich's approximation~\cite{kasai,zeld}, systematically derived in \cite{RZA1}.

We finally take advantage of this small-deviation picture to make a digression about the usefulness of the contrast density $\wt\Delta_\Sigma$ over the 
density contrast $\delta_\Sigma = \wt\Delta_\Sigma/(1 - \wt\Delta_\Sigma)$. Expressing relation~(\ref{eq:zeld}) in terms of $\delta_\Sigma$ yields
\beqn
	&& \ddot{\delta}_\Sigma + 2 \frac{\dot{a}_\Sigma}{a_\Sigma} \dot{\delta}_\Sigma - 4 \pi G \averageS\varrho \delta_\Sigma \nonumber \\
	&& + \delta_\Sigma \left(\ddot{\delta}_\Sigma + 2 \frac{\dot{a}_\Sigma}{a_\Sigma} \dot{\delta}_\Sigma - 8 \pi G \averageS\varrho \delta_\Sigma 
	- 4 \pi G \averageS\varrho \delta_\Sigma^2 \right) - 2 \dot{\delta}_\Sigma^2 = 0 \, . \label{eq:dens_cont}
\eeqn
\end{subequations}
This illustrates that the solution to the linear equation~(\ref{eq:zeld}) for $\wt\Delta_\Sigma$ substantially goes beyond that for $\delta_\Sigma$,
$\ddot{\delta}_\Sigma + 2 ({\dot{a}_\Sigma}/{a_\Sigma}) \dot{\delta}_\Sigma - 4 \pi G \averageS\varrho \delta_\Sigma = 0$. Hence, first-order Lagrangian 
(relativistic) deviations off a FLRW background already involve nonlinearities in the dependent variable $\delta_\Sigma$, which demonstrates the 
inherently nonlinear character of a Lagrangian perturbation approach. Remark at last that the density contrast coincides by construction with the density 
deviations $\delta\wt\varrho$ evaluated in the globally comoving metric; it is simply the relativistic extension of the density deviation field used 
in the standard Eulerian cosmological perturbation theory \cite{peebles} (see \cite{buchert:perturb,buchert:keymodel,ehlersbuchert} for other remarks on the Lagrangian 
picture versus the Eulerian one in Newtonian cosmology).

\section{Concluding remarks and outlook} \label{sec:conc}

We have generalized, in the present paper, the dust scalar perturbation scheme off a predefined background to deviations off 
a general background. The kinematical backreaction, which determines the global contribution of fluid 
inhomogeneities within a generic domain, is at the very core of this non-perturbative scheme: it not only influences the dynamics of a general 
background, as it is well known, but it also explicitly impacts on that of the deviation fields. 
Our long-term expectation from this improvement is to be able to describe large-scale structure formation uniquely from the existence of inhomogeneities 
and without the need to invoke a dark energy fundamental component.

Our scheme may be exactly solved either by considering a specific 3-metric, or by imposing local dynamical constraints, or by restricting 
attention to subclasses of backgrounds and then constraining the deviation fields. Another, and we think the most promising, strategy to solve the 
deviation equations would be to develop an iterative procedure. The reason why an iterative procedure takes better care, compared to a perturbative approach, 
of the nonlinear character of the proposed scheme is obvious: a perturbation point of view runs into 
contradiction due to the fact that the background (the zeroth-order solution) is generally modified by the kinematical backreaction (a second-order term). 
Hence, for situations where the backreaction term does not vanish, the notion of, e.g., first-order deviations off a general background would be 
ill-defined. An iterative point of view also entails methods that are known to numerical simulations, and we expect that the simplest application 
of the presented scheme is numerical in nature. This relativistic Lagrangian procedure would consist, for the zeroth-level, of taking the 
Zel'dovich approximation (\ref{eq:zeld}) and computing the first-level backreaction term. This first implementation has been depicted in 
\cite{buchert:focus} (see section 7.3). Reinjecting this first-level solution, at each step of time, into the background, perturbing this latter and 
solving the equations of the above corollaries, would then drive the second-level kinematical backreaction, and so on for the ensuing levels of iteration. 
This process should clearly be carried beyond the first levels of iteration in order that the inherently nonlinear character of structure formation, and its effects, 
shall be taken into account. Another promising approach would be to use this procedure in the framework of the gradient expansion treatment of inhomogeneities. 
Developed in \cite{croud,salope} and used in \cite{kolb:grad} to study backreaction, this technique---contrary to perturbation theory---seems to take 
better care of the impact of the small-scale nonlinear effects on the larger ones. 
A recent study \cite{enqvist:grad} has shown that, already for gradient expansion quantities of the fourth level, the effects of backreaction can grow up to 
10\% of the background. A realization of this iterative strategy is the subject of forthcoming work.

\subsection*{Acknowledgements:} 
{\small XR acknowledges support from the \'Ecole Doctorale de Lyon and the Claude Leon Foundation. 
We would like to thank C.\ Clarkson, H.\ L.\ Moute, N.\ Obadia, S.\ R\"as\"anen, A.\ Wiegand and D.\ Wiltshire 
for their comments and remarks, and C.\ Nayet for related discussions. 
This work was supported by Ò{\it F\'ed\'eration de Physique Andr\'e--Marie Amp\`ere Lyon}Ó, 
and was conducted within the Ò{\it Lyon Institute of Origins}Ó under grant ANR-10-LABX-66.}



{



\section*{References} 
\begin{thebibliography}{999}

\bibitem{alcub:fol}
Alcubierre M 2008 {\it Introduction to $3+1$ numerical relativity}
(Oxford: Oxford University Press)
%
\bibitem{bruni}
Bruni M, Matarrese S and Pantano O 1995 Dynamics of silent universes
{\it Astrophys.\ J.} {\bf 445}, 958
%
\bibitem{buchert89}
Buchert T 1989 A class of solutions in Newtonian cosmology and the pancake theory
{\it Astron.\ Astrophys.} {\bf 223} 9 
%
\bibitem{buchert92}
Buchert T 1992 Lagrangian theory of gravitational instability of Friedmann--Lema\^itre cosmologies and the `Zel'dovich approximation'
{\it Mon.\ Not.\ R.\ Astron.\ Soc.} {\bf 254} 729 
%
\bibitem{buchert:keymodel} 
Buchert T 1993 Lagrangian perturbation theory: a key model for large scale structure.
{\it Astron. Astrophys.} {\bf 267} L51
%
\bibitem{buchert:perturb}
Buchert T 1996 Lagrangian perturbation approach to the formation of large-scale structure
{\it Course CXXXII: Dark Matter in the Universe: Proc. of International School of Physics Enrico Fermi (Varenna 1995)}
ed S Bonometto, J Primack and A Provenzale (Amsterdam: Institute of Physics Publishing) p 543
arXiv:astro-ph/9509005
%
\bibitem{buchertehlers}
Buchert T and Ehlers J 1997 Averaging inhomogeneous Newtonian cosmologies
{\it Astron.\ Astrophys.} {\bf 320} 1
%
\bibitem{buchert:jgrg}
Buchert T 2000 On average properties of inhomogeneous cosmologies
{\it 9th JGRG Meeting (Hiroshima, 1999)} ed Y Eriguchi {\it et al} vol 9 p 306
arXiv:gr-qc/0001056
%
\bibitem{buchert:dust}
Buchert T 2000 On average properties of inhomogeneous fluids in general relativity: dust cosmologies
{\it Gen.\ Rel.\ Grav.} {\bf 32} 105
%
\bibitem{buchert:fluid}
Buchert T 2001 On average properties of inhomogeneous fluids in general relativity: perfect fluid cosmologies
{\it Gen.\ Rel.\ Grav.} {\bf 33} 1381
%
\bibitem{buchertdominguez:adhesive}
Buchert T and Dom\' \i nguez A 2005 Adhesive gravitational clustering
{\it Astron.\ Astrophys.} {\bf 438} 443 
%
\bibitem{buchert:static}
Buchert T 2006 On globally static and stationary cosmologies with or without a cosmological constant and the dark energy problem
{\it Class.\ Quantum Grav.} {\bf 23} 817
%
\bibitem{buchert:morphon}
Buchert T, Larena J and Alimi J M 2006 Correspondence between kinematical backreaction and scalar field cosmologies---the `morphon field'
{\it Class.\ Quantum Grav.} {\bf 23} 6379 
%
\bibitem{buchert:nonperturbative}
Buchert T 2006 The non-perturbative regime of cosmic structure formation
{\it Astron.\ Astrophys.} {\bf 454} 415 
%
\bibitem{buchert:review}
Buchert T 2008 Dark energy from structure: a status report
{\it Gen.\ Rel.\ Grav.} {\bf 40} 467
%
\bibitem{buchertcarfora}
Buchert T and Carfora M 2008 On the curvature of the present day universe
{\it Class.\ Quantum Grav.} {\bf 25} 195001 
%
\bibitem{buchert:estim}
Buchert T, Ellis G F R and Van Elst H 2009 Geometrical order-of-magnitude estimates for spatial curvature in realistic models of the universe
{\it Gen.\ Rel.\ Grav.} {\bf 41} 2017
%
\bibitem{buchert:inf}
Buchert T and Obadia N 2011 Effective inhomogeneous inflation: curvature inhomogeneities of the Einstein vacuum
{\it Class.\ Quantum Grav.} {\bf 28} 162002 
%
\bibitem{buchert:focus}
Buchert T 2011 Toward physical cosmology: focus on inhomogeneous geometry and its non-perturbative effects
{\it Class.\ Quantum Grav.} {\bf 28} 164007 (Focus section `Inhomogeneous cosmological models and averaging in cosmology)
%
\bibitem{buchertrasanen}
Buchert T and R\"as\"anen S 2012 Backreaction in late-time cosmology
{\it Annu. Rev. Nucl. Part. Sci.} {\bf 62} 57 at press
arXiv:1112.5335
%
\bibitem{RZA1}
Buchert T and Ostermann M 2012 
Lagrangian theory of structure formation in relativistic cosmology: I Lagrangian framework and definition of a non-perturbative approximation
arXiv:1203.6263 
%
\bibitem{RZA2}
Buchert T, Nayet C and Wiegand A Lagrangian theory of structure formation in relativistic cosmology: II A generic evolution model for average characteristics
(in preparation)
%
\bibitem{clarkson:pert}
Clarkson C, Clifton T and February S 2009 Perturbation theory in Lema\^itre--Tolman--Bondi cosmology
{\it J. Cosmol. Astropart. Phys.} JCAP06(2009)025
%
\bibitem{clarkson:inho_conc}
Clarkson C and Maartens R 2010 Inhomogeneity and the foundations of concordance cosmology
{\it Class.\ Quantum Grav.}  {\bf 27} 124008
%
\bibitem{clarkson:review} 
Clarkson C, Ellis G F R, Larena J and Umeh O 2011
Does the growth of structure affect our dynamical models of the Universe? The averaging, backreaction, and fitting problems in cosmology
{\it Rep. Prog. Phys.} {\bf 74} 112901
%
\bibitem{croud}
Croudace K M, Parry J, Salopek D S and Stewart J M 1994 Applying the Zel'dovich approximation to general relativity
{\it Astrophys.\ J.} {\bf 423} 22
%
\bibitem{ehlersbuchert}
Ehlers J and Buchert T 1997
Newtonian cosmology in Lagrangian formulation: foundations and perturbation theory
{\it Gen. Rel. Grav.} {\bf 29} 733
%
\bibitem{ellis:focus}
Ellis G F R 2011 Inhomogeneity effects in cosmology
{\it Class.\ Quantum Grav.} {\bf 28} 164001 (Focus section `Inhomogeneous cosmological models and averaging in cosmology')
%
\bibitem{LTB:review}
Enqvist K 2008 Lema\^itre--Tolman--Bondi model and accelerating expansion
{\it Gen.\ Rel.\ Grav.} {\bf 40} 451
%
\bibitem{enqvist:grad}
Enqvist K, Hotchkissa S and Rigopoulos G 2012 A gradient expansion for cosmological backreaction.
{\it J. Cosmol. Astropart. Phys.} JCAP03(2012)026
%
\bibitem{gasp}
Gasperini M, Marozzi G and Veneziano G 2010
A covariant and gauge invariant formulation of the cosmological `backreaction'
{\it J. Cosmol. Astropart. Phys.} JCAP02(2010)009
%
\bibitem{giesel:pert}
Giesel K, Hofmann S, Thiemann T and Winkler O 2010 Manifestly gauge-invariant general relativistic perturbation theory: I. Foundations
{\it Class.\ Quantum Grav.} {\bf 27} 055005
%
\bibitem{gourg:fol}
Gourgoulhon E 2012 
{\it $3+1$ Formalism in General Relativity. Bases of Numerical Relativity (Lecture Notes in Physics} vol 846) (Berlin: Springer)
%
\bibitem{kasai}
Kasai M 1995 Tetrad-based perturbative approach to inhomogeneous universes: a general relativistic version of the Zel'dovich approximation
{\it Phys.\ Rev.\ D} {\bf 52} 5605
%
\bibitem{kofman}
Kofman L and Pogosyan D 1995 Equations of gravitational instability are non-local
{\it Astrophys.\ J.} {\bf 442} 30 
%
\bibitem{kolb:grad}
Kolb E W, Matarrese S and Riotto A 2006 On cosmic acceleration without dark energy
{\it New J.\ Phys.} {\bf 8} 322
%
\bibitem{kolb:voids}
Kolb E W, Marra V and Matarrese S 2008 Description of our cosmological spacetime as a perturbed conformal Newtonian metric
and implications for the backreaction proposal for the accelerating universe
{\it Phys.\ Rev.\ D} {\bf 78} 103002
%
\bibitem{kolb:backgrounds}
Kolb E W, Marra V and Matarrese S 2010 Cosmological background solutions and cosmological backreactions
{\it Gen.\ Rel.\ Grav.} {\bf 42} 1399
%
\bibitem{kolb:focus}
Kolb E W 2011 Backreaction of inhomogeneities can mimic dark energy
{\it Class.\ Quantum Grav.} {\bf 28} 164009  (Focus section `Inhomogeneous cosmological models and averaging in cosmology')
%
\bibitem{matarrese:pert}
Matarrese S and Mohayaee R 2002 The growth of structure in the intergalactic medium
{\it Mon.\ Not.\ R.\ Astron.\ Soc.} {\bf 329} 37
%
\bibitem{matsumoto}
Matsumoto J 2011 On the necessity of the revisions for the cosmological matter perturbations from the general relativity
{\it Phys.\ Rev.\ D} {\bf 83} 124040 
%
\bibitem{morita}
Morita M, Nakamura K and Kasai M 1998 Relativistic Zel'dovich approximation in spherically symmetric model
{\it Phys.\ Rev.\ D} {\bf 57} 6094 
%
\bibitem{abramo1}
Mukhanov V F, Abramo L R W and Brandenberger R H 1997 Backreaction problem for cosmological perturbations
{\it Phys.\ Rev.\ Lett.} {\bf 78} 1624
%
\bibitem{nakamura}
Nakamura K 2003 Gauge invariant variables in two-parameter nonlinear perturbations
{\it Prog.\ Theor.\ Phys.} {\bf 110} 723
%
\bibitem{peebles}
Peebles P J E 1980 
{\it The Large Scale Structure of the Universe}
(Princeton, NJ: Princeton University Press)
%
\bibitem{rasanen:de}
R\"as\"anen S 2004 Dark energy from backreaction
{\it J. Cosmol. Astropart. Phys.} JCAP02(2004)003
%
\bibitem{rasanen:acceleration}
R\"as\"anen S 2006 Accelerated expansion from structure formation
{\it J. Cosmol. Astropart. Phys.} JCAP11(2006)003 
%
\bibitem{rasanen:perturbation}
R\"as\"anen S 2010 Applicability of the linearly perturbed FRW metric and Newtonian cosmology
{\it Phys. Rev. D} {\bf 81} 103512 
%
\bibitem{rasanenFOCUS} 
R\"as\"anen S 2011 Backreaction: directions of progress
{\it Class.\ Quantum Grav.} {\bf 28} 164008 (Focus section `Inhomogeneous cosmological models and averaging in cosmology')
%
\bibitem{roy:chap}
Roy X and Buchert T 2010 Chaplygin gas and effective description of inhomogeneous universe models in general relativity
{\it Class.\ Quantum Grav.} {\bf 27} 175013
%
\bibitem{roy:ps}
Roy X, Buchert T, Carloni S and Obadia N 2011 Global gravitational instability of FLRW backgrounds---interpreting the dark sectors
{\it Class.\ Quantum Grav.} {\bf 28} 165004
%
\bibitem{russ1}
Russ H, Morita M, Kasai M and B\"orner G 1996 The Zel'dovich-type approximation for an inhomogeneous universe in general relativity: second-order solutions
{\it Phys.\ Rev.\ D} {\bf 53} 6881
%
\bibitem{russ2}
Russ H, Soffel M H, Kasai M and B\"orner G 1997 Age of the universe: influence of the inhomogeneities on the global expansion-factor
{\it Phys.\ Rev.\ D} {\bf 56} 2044
%
\bibitem{salope}
Salopek D S, Stewart J M and Croudace K M 1994 The Zel'dovich approximation and the relativistic Hamilton--Jacobi equation
{\it Mon.\ Not.\ R.\ Astron.\ Soc.} {\bf 271} 1005
%
\bibitem{smarr:fol}
Smarr L and York J W 1978 Kinematical conditions in the construction of spacetime
{\it Phys.\ Rev.\ D} {\bf 17} 2529 
%
\bibitem{schur}
Schur I 1905 
Modern algebraic theories {\it Neue Begr\"undung der Theorie der Gruppencharaktere Sitzungsber}
vol 406 (Berlin: K Preuss Akad. Wiss.) chapter 11 Translated by L E Dikson and reprinted in 1926 (Chicago: Sanborn and Co.)
%
\bibitem{sussman:ltb}
Sussman R A 2011 Back-reaction and effective acceleration in generic LTB dust models
{\it Class.\ Quantum Grav.} {\bf 28} 235002
%
\bibitem{labini:focus}
Sylos Labini F 2011 Inhomogeneities in the universe
{\it Class.\ Quantum Grav.} {\bf 28} 164003 (Focus section `Inhomogeneous cosmological models and averaging in cosmology')
%
\bibitem{momoko1}
Tatekawa T, Suda M, Maeda K and Kubotani H 2001 Inhomogeneities in Newtonian cosmology and its backreaction to the evolution of the universe
arXiv:astro-ph/0109501
%
\bibitem{momoko2}
Tatekawa T, Suda M, Maeda K, Morita M and Anzai H 2002 Perturbation theory in Lagrangian hydrodynamics for a cosmological fluid with velocity dispersion.
{\it Phys.\ Rev.\ D} {\bf 66} 064014
%
\bibitem{trump}
Tr\"umper M 1967
Bemerkungen \"uber scherungsfreie Str\"omungen gravitierender Gase 
{\it Zeits.\ f.\ Astrophys.} 66 215
%
\bibitem{multiscale}
Wiegand A and Buchert T 2010 Multiscale cosmology and structure-emerging dark energy: a plausibility analysis
{\it Phys.\ Rev.\ D.} {\bf 82} 023523
%
\bibitem{wiltshire:focus} 
Wiltshire D L 2011 What is dust? -- Physical foundations of the averaging problem in cosmology
{\it Class.\ Quantum Grav.} {\bf 28} 164006 (Focus section `Inhomogeneous cosmological models and averaging in cosmology')

\bibitem{wiltshire:cosmic} 
Wiltshire D L 2007 Cosmic clocks, cosmic variance and cosmic averages
{\it New J.\ Phys.} {\bf 9} 377
%
\bibitem{wiltshire:average} 
Wiltshire D L 2009 Average observational quantities in the timescape cosmology
{\it Phys.\ Rev.\ D} {\bf 80} 123512
%
\bibitem{zeld}
Zel'dovich Ya B 1970 Gravitational instability: an approximate theory for large density perturbations
{\it Astron.\ Astrophys.} {\bf 5} 84 
%
\bibitem{zhao}
Zhao X and Mathews G J 2011 Effects of structure formation on the expansion rate of the universe: an estimate from numerical simulations
{\it Phys.\ Rev.\ D} {\bf 83} 023524
%
\end{thebibliography}
\end{document}